\documentstyle[prl,aps]{revtex}
\begin{document}
\draft
\title{On the Polarization Dependence of\\ Electromagnetic Vertex for Proton}
\author{C.B. Yang and X. Cai}
\address{Institute of Particle Physics, Hua-Zhong Normal University,
Wuhan 430070, China}
\maketitle

\begin{abstract}
The polarization dependence of the electromagnetic vertex for polarized
proton is discussed. Such a dependence is shown to be a result of
gluon polarization and gluon interactions with the constituents within
proton. One more form factor is introduced to give such a dependence in
the electromagnetic vertex for proton. Due the polarization dependence of
the vertex, a nonzero single-spin asymmetry for unpolarized lepton
scattering off transversely polarized proton is predicted.
\end{abstract}



One of the most critical challenges facing the modern particle physicists
is the understanding of nucleon structure.
For a long time the theory of hadron structure remained
comfortably at the level of the naive parton model, because a vast of
unpolarized experimental data can be explained within that model.
Experiments on polarized deep inelastic lepton-nucleon scattering
started in the middle 70s [1]. In 1988, the EMC Collaboration [2] measured
the first moment of the proton spin structure function, $\Gamma^p_1
\equiv\int^1_0 g^p_1(x)\,{\rm d}\,x$, and their results differed significantly
from the naive theoretical predictions and known as the EMC effect.
The EMC effect motivated intensive further experimental and theoretical
activities. During the period of 1988-1993, theorists tried hard
to resolve the proton spin enigma and seek explanation for the EMC effect,
based on assuming the validity of the EMC data at small $x \ (0.01<x<0.1)$
and of the extrapolation procedure to the unmeasured small $x$ region
$(x<0.01)$. The discussions on proton spin have been continuing up to now.
Two successful first-principles calculations [3,4] of the quark spin contents
based on lattice QCD were published very recently. The calculations
revealed that the sea-quark polarization arises from the disconnected
insertion and is empirically SU(3)-flavor symmetric. This suggests that
the conventional practice of decomposing $\Delta q$ into valence and sea
components is not complete, and the effect of ``cloud'' quarks
should be taken into account. These calculations also show that the
non-perturbative effects play an important role in spin phenomena.
On the experimental side, deep-inelastic scattering experiments of
polarized leptons off polarized nucleon were carried out
recently at CERN [5] and SLAC [6]. The HERMES experiment will
work at the $ep$ collider HERA at DESY [7], which will give access
to a so far unexplored kinematic region of lepton-nucleon scattering
with rich physical potential. It is now well known that the physics
of the polarized hadron-hadron interactions is very different from
that of the unpolarized. 'Polarization data has often been the graveyard of
fashionable theories. If theorists had their way they might well ban such
measurements altogether out of self-protection.' [8]. Indeed, the EMC effect
on the spin dependent structure functions [2] has destroyed
our naive expectations [9], and the role played by proton spin seems to
be more important than that once assumed.

In this Letter we come down to the very starting point of current
theories for lepton-nucleon scatterings and try to seek whether there are
some contents on the spin dependent effects not included in
traditional theories yet. We study the elastic scattering of
unpolarized lepton off polarized proton and seek
for the parameterization for the electromagnetic vertex for
proton. It is found that the traditional expression for the electromagnetic
vertex for proton should be modified into the form containing a polarization
dependent term and an additional form factor should be introduced to describe
the polarization dependence.

Consider elastic lepton scattering off proton. Denote $m$ the
lepton mass, $k\ (k^{\prime})$ the initial (final) lepton four-momentum
and $M$ the proton mass, $P$ and $S$ the initial proton four-momentum
and spin four-vector. In current theories, the electromagnetic vertex for
proton in elastic lepton-proton scatterings
was assumed to be independent of the initial proton's spin orientation
and was assumed to take a simple form as
\begin{equation}
J_\mu=e\bar{u}(P^\prime)\,\Gamma^{({\rm trad})}_\mu\,u(P)\ \ ,
\end{equation}

\noindent with traditional expression for the electromagnetic vertex
\begin{equation}
\Gamma^{({\rm trad})}_\mu=F_1(q^2)\gamma_\mu+
{\kappa\over 2M}F_2(q^2) i\sigma_{\mu\nu}q^\nu\ \ ,
\end{equation}

\noindent
and Dirac spinors $u(P)$ and $\bar{u}(P^\prime)$ are the same as those
for free point-like particle with mass $M$, $\kappa$ is the anomalous
part of the magnetic moment in units of the Bohr magneton and scalar
functions $F_1(q^2)$ and $F_2(q^2)$ are called form factors.

It is worthy to recall some facts:
(i) The  proton transition current of Eq.(1) was derived at a time
when proton was regarded as elementary point-like particle;
(ii) At that time the interaction between nucleons was regarded as
a copy of electromagnetics,  with massive $\pi$ in place of
massless photon; (iii) Spin effects were presumed at that time not
so important as today, and one was usually satisfied with the spin
averaged results; (iv) No polarized experiment was done
in early years. Though theories based on Eqs. (1) and (2)
can explain lots of unpolarized experiments, the validity of Eq. (2)
for $\Gamma^{({\rm trad})}_\mu$ in polarized processes had not been actually
verified after the appearance of polarized lepton-nucleon experiments.

It is also fruitful here to recall that the transition current for proton
(Eq. (1)) was assumed in current theories to be the same form as that
for an electron in external electromagnetic field with quantum fluctuation
considered. The only difference between them is that values of the
form factors  are different. It should be pointed out
that the terminology `form factor' implies not the compositeness
of the considered particle but merely the existence of quantum
fluctuation. So does the terminology `structure function'.
Now we know that proton is composed of quarks and gluons with interaction
inside governed by the QCD. Thus it is clear that proton is very different
from electron in: (i) Proton is a composite extended object while
electron has no constituents; (ii) Interaction inside proton is mediated
by exchanging colored gluons with nonabelian quantum fluctuations,
while the quantum fluctuations for electron are mediated by the exchange of
abelian photon. Keeping all those differences mentioned above in
mind, one may ask why the electromagnetic current for proton
should not take a different form from that for electron.

To answer this question let's investigate the photon interaction with
polarized proton, taking into account the effects of strong interactions
inside proton.

First of all, it is inevitable to accept the existence of hadronic current
inside a polarized hadron as long as one accepts the concept of an extended
hadron with a hadronic matter density in it, as pointed out in Ref. [10].
The hadronic current inside a polarized hadron can be understood when one
notices that valence quarks inside hadron are relativistic Dirac particles
moving in a confining field produced by other quarks. The orbital motion
of such a relativistically moving Dirac particle is always involved $-$
also when such particles are in ground states. Thus the orbiting quarks
can induce the hadronic current. A particular feature of this hadronic
current is that it depends on the polarization of the hadron. It is without
any doubt that such a hadronic current in a polarized hadron will induce
observable effects besides those mentioned in Ref. [10].

On the other hand, the gluon fields inside a polarized hadron must also be
polarized because they are generated by the polarization dependent hadronic
current. This can immediately be seen from
the classical Yang-Mills equation for strong interactions. Very recently,
Ji [11] has obtained gauge-invariant expressions for the angular momenta
for quarks and gluons inside a proton, both of which depend on proton's
polarization, indicating quark's orbital motion and gluon's polarization.
In fact, the polarization of gluons had been noticed several year ago [12]
theoretically and experimental evidences [2] show that gluons might
contribute almost total of proton's spin. Thus the net polarization of gluons
should be along that of proton. Many authors have tried to account for the EMC
effect in terms of anomalous gluon contribution [13-16].

Now one can evaluate the influence of the polarized gluon field on
quark's wavefunction. From Dirac equation, the wavefunction of a quark
inside polarized hadron can be formally written as
\begin{equation}
\Psi={1\over 1-(\hat{p}-m)g\hat{B}}\,\Psi_0\ ,
\end{equation}

\noindent where $\hat{A}\equiv\gamma^\mu A_\mu$, $p$ the four momentum of
the quark, $B_\mu$ the gluon fields acting on the quark, $g$ the coupling
constant, and $\Psi_0$ Dirac spinor satisfying $(\hat{p}-m)\Psi_0=0$.

From the view point of field theory,
the photon-quark interaction can be described by the Lagrangian
\begin{equation}
{\cal L}={\cal L}_0+{\cal L}_{\rm strong}+{\cal L}_{\rm em}\ .
\end{equation}

\noindent
With this Lagrangian, one can prove that the transition current for the
quark is [17]
\begin{equation}
J^q_\mu=e_q{\overline \Psi}(p^\prime,s^\prime)\, \Gamma^{(q)}_\mu\, \Psi(p,s)\ \ ,
\end{equation}

\noindent
with the form of $\Gamma^{(q)}_\mu$ the same as that given in Eq. (2).
It must be emphasized that both $\Psi(p,s)$ and ${\overline \Psi}
(p^\prime,s^\prime)$ in last equation
are eigenstates of Hamiltonian $H_0+H_{\rm strong}$, i.e., they are not
free states of the quark but contain information about
the gluon interactions inside polarized proton.
One has seen from above discussions that $B_\mu$ depends on the polarization
of proton, so that it is natural that the transition current for the quark
inside polarized proton depends not only on the quark's spin $s$ but also on
proton's polarization $S$. Last equation can be expressed in terms of states
$\Psi_0$ as follows
\begin{equation}
J^q_\mu=e_q{\overline \Psi}_0(p^\prime, s^\prime)\,\Gamma^q_\mu(p,p^\prime,
S)\,\Psi_0(p,s)\ \ ,
\end{equation}

\noindent with $\Gamma^q_\mu$ depending on proton's polarization vector
$S$.

The electromagnetic current of proton is the sum of
those of quarks. Assuming the electromagnetic vertex for quarks
is simply $e_q\gamma_\mu$, the electromagnetic vertex for elastic
scattering between proton and photon should be of the form
shown in Eq. (2). From above discussions, the vertex for photon-quark
interaction depends on proton polarization, thus one has to admit that
the vertex for proton-photon interaction should depend on the polarization
$S$ of initial proton. So the electromagnetic current for proton in
elastic lepton-proton scattering is
\begin{equation}
J_\mu=e{\overline u}(P^\prime, S^\prime)\,\Gamma_\mu(P,P^\prime,
S)\,u(P,S)\ \ ,
\end{equation}

\noindent
and $\Gamma_\mu$ should depend on proton polarization.
Because of the never mentioned polarization
dependence of $\Gamma_\mu$, some modifications to the traditional expression
for the electromagnetic vertex $\Gamma^{({\rm trad})}_\mu$ for proton  must
be made. Physically it demands Lorentz covariance and current conservation,
i.e. $J_\mu$ should be a true four-vector and satisfy $q^\mu J_\mu=0$.
The most general form of $\Gamma_\mu$, which is composed
of $\gamma_\mu, P_\mu, q_\mu$ and $S_\mu$ and makes
all those physical restrictions satisfied, can be written as
\begin{equation}
\Gamma_\mu=F_1\gamma_\mu+{F_2\over 2M}
i\sigma_{\mu\nu}q^\nu+{F_3\over 2M}
\varepsilon_{\mu\nu\lambda\tau}\,\gamma^\nu S^\lambda q^\tau\ \ ,
\end{equation}

\noindent where, different from traditional ones, form factors $F_1, F_2$
and $F_3$ are real functions of invariant variables $q^2$ and $(S\cdot q)^2$.
The newly introduced $F_3$ term offers the polarization dependence of
proton structure, which can be attributed to the contribution of polarized
gluons. Clearly, the additional term in last equation conserves the current
and has proper parity because of the product of two pseudo-tensors
$\varepsilon^{\mu\nu\lambda\tau}$ and $S_\lambda$. Furthermore, the additional
term has the same time reversal properties as the first two terms, which
guarantees the $T$-invariance of the electromagnetic current for
polarized proton.

Because of the polarization dependent term in the electromagnetic
vertex for proton, new phenomena for elastic lepton-proton scatterings
may exist. It is straightforward to show the symmetric part (or real part)
of the hadronic tensor for proton to be as following
\begin{eqnarray}
W_{\rm symm}^{\mu\nu} &=& \left(8F^2_1-4F^2_2\left(1+{q^2\over 2M^2}\right)
\right) \left(P^\mu-{P\cdot q\over q^2}q^\mu\right)\left(P^\nu-{P\cdot
q\over q^2}q^\nu\right)\nonumber\\
&+& 2(F_1+F_2)^2q^2\left(g^{\mu\nu}-{q^\mu q^\nu\over q^2}\right)
+{2F^2_3\over M^2}\varepsilon^{\mu\xi\eta\theta}\varepsilon^{\nu\rho
\lambda\tau}S_\eta S_\lambda q_\theta q_\tau\left(P_\xi P_\rho+{q^2\over 4}
g_{\xi\rho}\right)\nonumber\\
&+& {4F_1F_3\over M}S_\lambda q_\tau P_\rho\left[\varepsilon^{\mu\rho\lambda
\tau}\left(P^\nu-{P\cdot q\over q^2}q^\nu\right)+\mu\leftrightarrow\nu\right]
\ .
\end{eqnarray}

\noindent Clearly, this hadronic tensor satisfies Eq. (42) in Ref. [18] and
thus is $T$-invariant. When proton spin is reversed, terms in the first two
lines will not change sign, but terms in the third sign will. If one considers
unpolarized elastic lepton-proton scattering, terms in the third line
contribute zero, and terms in the second line of last equation
can be re-parameterized into the same forms as terms in the first line.
Due to the presence of polarization dependent terms in the hadronic tensor
for proton $W^{\mu\nu}_{\rm symm}$, one can predict a single-spin asymmetry
for unpolarized lepton scattering off polarized proton. Let $L_{\mu\nu}$
be the lepton tensor in the process and $S$ be the polarization of initial
proton. If the initial proton is rest, the single-spin asymmetry can be
expressed as
\begin{eqnarray}
A & =& {{\rm d}\sigma(S)-{\rm d}\sigma(-S)\over
{\rm d}\sigma(S)+{\rm d}\sigma(-S)}\nonumber\\
&= & {16F_1F_3(k_0-q_0){\bf k}\cdot ({\bf q}\times {\bf S})\over
L_{\mu\nu}\,\left(W^{\mu\nu}_{\rm symm}(S)+W^{\mu\nu}_{\rm symm}(-S)\right)}\ .
\end{eqnarray}

It is obvious that a zero value is obtained for the asymmetry $A$
if the proton is longitudinally polarized. This result is not
surprising because of the symmetric space about the polarization axis.
Such an asymmetry is nonzero in the weak interactions because of
parity and $T$-invariance violations in weak interactions
was used many years ago to detect the weak interaction effects in the
scatterings. Our result confirms that there exists no longitudinal
polarization asymmetry for electromagnetic interactions because of $P$- and
$T$-invariances. For the case of transversely polarized proton,
however, the asymmetric might exist. Similar asymmetry in hadron-hadron
scatterings has been reported, which is very different from zero and
larger than predicted from the perturbative QCD. In this Letter,
we predict the existence of nonzero asymmetry for the case of
unpolarized lepton scattering off transversely polarized proton.
We show that this asymmetry is caused due to the interactions with
constitute quarks inside polarized proton of polarized gluon field.
It is due to such interactions that the transverse momentum distribution
of quarks inside proton depends on proton polarization whose effects
on single-spin symmetry in DIS has just been investigated in Ref. [19].

Unfortunately no experimental data on asymmetry of single polarized
lepton-proton scatterings are available yet, owing to the fact that
all current theories predicted a zero result. To check the polarization
dependence of the electromagnetic vertex for proton the measurement of
single-spin asymmetry $A$ defined above is necessary
and sufficient which needs to be suggested at CERN and DESY.
Recently the SMC collaboration has made a measurement of longitudinally
polarized muons scattered from transversely polarized proton [5]. Their
techniques can be used to test the single-spin asymmetry predicted
in this Letter.

As a summary we discussed the polarization dependence of the electromagnetic
vertex, $\Gamma_\mu$, for proton in elastic lepton-proton
scatterings. Our arguments are based on the consideration of the
gluon polarization and its interactions with constituents within the proton.
The polarization dependence of the vertex is given by the third form factor,
$F_3$, suggested in this Letter. Due to this new form factor for polarized
proton, there should exist nonzero single-spin asymmetry for unpolarized
lepton scattering from transversely polarized proton. Such an asymmetry
needs to be verified experimentally. If a positive result can be obtained
from future experiments, great changes will happen in both theoretical and
experimental studies of elastic and deep inelastic scattering processes.

This work was supported in part by the NNSF and the SECF in China.
The authors thank Prof. T. Meng for helpful discussions.

\end{document}